\documentstyle[prb,twocolumn,aps]{revtex}
\input{epsf}
\begin{document}

\title{A new method for a local study of nonlinear microwave
properties of superconductors}

\author{Evgeny E.Pestov, Yury N.  Nozdrin, \\ Vladislav V.Kurin}

\address{Institute for Physics of Microstructures, Russian
Academy of Sciences, Nizhny Novgorod, GSP-105, 603600, RUSSIA\\
}
 \maketitle
\begin{abstract}
We report a set of experimental data on local third-harmonic
generation at microwave frequencies (0.5 GHz) in $YBa_2Cu_3O_7$
and Nb films.  For local investigations of the nonlinear
response a probe with inductive coupling was elaborated. The map
of the nonlinear microwave response of a $YBa_2Cu_3O_7$ thin film
is plotted below $T_c$ with high resolution. The third-harmonic
power is measured as a function of temperature, input power and
dc magnetic field at some areas of the film. The correlation
between the depinning current density ${J_p}$ and
the nonlinear microwave response is also demonstrated.
\end{abstract}

HTS films find extensive application in passive
microwave devices such as transmission lines, antennas and
filters \cite{filter}, owing to a low surface resistance $R_s$
of HTS films, which ensures low losses at low power levels.  At
higher power surface resistance $R_s$ increases, due to
nonlinear properties of these films which leads to higher losses
and a shift of the resonance frequency, thus limiting the
applicability range for these films. The nonlinearity of surface
resistance $R_s$ is generally associated with the
Ginzburg-Landau nonlinearity \cite{gl}, thermal nonlinearity
\cite{therm}, hysteretic losses \cite{gis}, and the Josephson
nonlinearity\cite{gos}.  Yet, despite the numerous experiments
carried out in this field, the origin of a nonlinear microwave
response of superconductor has not been fully understood thus
far.  Therefore, investigation of nonlinearity is equally
important in terms of gaining an insight into the fundamental
properties of HTS films and towards applications in
superconducting electronics.

    In this work we propose a new technique for measuring a
nonlinear local microwave response of a superconducting film basing
on developed a near-field inductive coupling probe. Using this
method, we have plotted a map of the nonlinear local response of
a HTS film at a temperature below $T_c$, and also measured the
third-harmonic power as a function of input power, temperature,
and applied dc magnetic field for HTS and Nb films.

    A wide use in measurements of the nonlinear properties of
superconducting films lately has been made of a resonator
technique which allows to produce  rf magnetic
fields that are close in value to the characteristic rf magnetic
fields of nonlinearity. This is achieved by using a stripline
resonator \cite{gos} or a cavity resonator with a sample placed
inside\cite{lev}.  Characteristic values for the current
density or the magnetic field of film nonlinearity are
determined either from the power dependence of the
third-harmonic power \cite{wilker} or from measurements of
surface impedance which is related to microwave losses \cite{gos}.
Note that  the averaged nonlinear characteristics of a
microwave device are measured in this case.  Local measurements
of surface resistance $R_s$ in the microwave range are aided by
near-field microscopes in wide use currently.  Essentially, the
idea of near-field microscopy is in localization of a magnetic
or an electric field near probe on scales much less than a
wavelength.  A variety of probe designs conventionally used for
local investigations includes a circular aperture\cite{nature},
open-ended coaxial cable\cite{merz}, small loop\cite{cummings}, etc.

Here we present an original near-field probe used for
measuring a local nonlinear response of superconducting films,
that has been designed with due regard for the earlier developed
methods and approaches. A block diagram of the probe is shown in
Fig.1. The probe is essentially a 2 mm long 50 $\mu$m diameter
wire connecting the outer and the inner conductors of a
coaxial cable. Reflection of a microwave  signal from such a
probe gives rise to a high current flow in the wire
because the probe impedance is much less than the wave
impedance of the coaxial cable. The current induces a fairly
strong quasistatic magnetic field localized on a scale of order
of the probe diameter. The nonlinear properties of a
superconducting film are responsible for generation of higher
harmonics which are picked up by the same probe. The incident
wave frequency is 472 MHz, and the nonlinear response is
measured at the third harmonic frequency of 1.42 GHz. To avoid
contact effects preventing observation of a superconducting film
nonlinearity, a 10 $\mu$m thick teflon film is placed between the
probe and the sample.  Note that for the dimensions and the geometry of
the probe studied here the maximum of power of about 100 mW
produces a maximum current density $J_{rf}$ in the 100 nm film
of about $10^6 A/{cm}^2$.

    In this work we did an experimental study of 30-100 nm
thick $YBa_2Cu_3O_7$ films magnetron sputtered on a $GaNdO_3$
substrate.  The films quality was quite high (critical
current density of $\sim 10^6 A/cm^2$ ).  We also investigated
Nb films of 30 nm thickness.

    Fig.2 shows a temperature dependence of the third-harmonic
signal at different levels of input power, which features a
nonlinearity peak below $T_c$. It should be noted that
nonlinearity maxima near $T_c$ were observed in a number of
works \cite{gl,hampel}.  In \cite{hampel} the nonlinearity peak
was shown to appear by penetration of a magnetic flux through
the film edges.  In our case the probe was placed in a film
center, but a sharpest peak remained. For a qualitative
analysis of the temperature dependence of nonlinear response,
we used  the measurements of the temperature dependences of the
depinning current density $J_p$, found from
measurements of the residual magnetization produced in a film by
an external uniform magnetic field, the current density of vortex
penetration $J_c$ (which corresponds to the Ginzburg-Landau pair-breaking
current density $J_{GL}$ for perfect superconductor )
and the resistivity $\rho$, kindly provided by the authors of
\cite{meln}.  By comparing these dependences we found out that the
temperature of the peak $T_{max}$, correlates with that at which the
depinning current density $J_p$ disappears, and the temperature of
nonlinearity vanishing corresponds to the off-set temperature
$\rho(T)$.  The correlation between the nonlinear microwave
properties and the depinning current density indicates that the
nonlinearity observed at temperatures close to $T_c$ is of a
vortex origin.

    In Fig.3 the third-harmonic power
$P_{3\omega}(P_{\omega}$) is shown as a function of input
power on a log-log scale for $YBa_2Cu_3O_7$ and Nb films.
The data are readily approximated by the power law
$P_{3\omega}\sim A {P_{\omega}}^n$ . At temperatures close
to $T_c$ the HTS films exhibit a deviation from the exponent n =
3 (which is characteristic of an ordinary cubic nonlinearity
described by the Ginzburg-Landau equations), while Nb films
feature a marked power threshold. The exponent deviation from n
= 3 for HTS films occurs through saturation of the power
dependence of the third-harmonic signal at high  input powers or
at temperatures close to $T_c$ \cite{hampel}.

        The third-harmonic power $P_{3\omega}(H_{dc})$  as a function of
a dc magnetic field $H_{dc}$ at a temperature near the
nonlinearity peak and at T = 77 K is shown in Fig.4. The behavior
of $P_{3\omega}(H_{dc})$ differs qualitatively at liquid nitrogen
temperature and temperatures close to $T_c$.  At T = 77 K there
is a rise in the third-harmonic value and a fairly strong
hysteresis is observed, whereas in the vicinity of $T_c$
irreversibility disappears and an increase in the field causes
suppression of nonlinearity. A strong dependence on an external
magnetic field is also evidence of the vortex origin of the
observed nonlinearity; it may be connected with a decrease in
the depinning  current density $J_p$ at higher
temperatures (Fig.2).

    The method developed was used to map a nonlinear local
microwave response for $YBa_2Cu_3O_7$  at liquid nitrogen
temperature, as shown in Fig.5. We have chosen a positioning
system such that it would allow the probe to be moved at a 125
$\mu$m step in the direction of the x and y axes. Fig.5
demonstrates a nonuniform distribution of the nonlinear response
across the film surface, which depends on the inhomogeneity of
the critical current density in the sample. Note also that
nonlinearity increases in areas lying closer to the sample edge,
when the probe is parallel to the film boundary. This effect can
be explained by an increasing density of the current excited in
the film.

Although a complete quantitative analysis of the experimental data is
impossible currently, some qualitative considerations seem plausible
enough. Estimated value of the highest current density
$J_{rf}\sim10^5-10^6 A/{cm}^2$ in the film is higher or of the order of
the current density of vortex penetration $J_c$ and it is naturally to
consider that the nonlinear response is due to creation of vortices by
microwave field. At the same time the relation between the temperature
dependences of the nonlinearity and the depinning current density $J_p$
(Fig.2), and disappearance of irreversibility in the dc magnetic field
dependence of the third-harmonic power (Fig.4) demonstrate the
substantial role of thermal fluctuations in the vortex response. The
nonlinear response of low-temperature superconductors, unlike in the HTS
case, demonstrates a power threshold likely to be related to the onset of
vortex creation by the microwave field.

In summary, a new method for local investigation of the
nonlinear microwave properties of superconducting films has been
developed and used for mapping of a nonlinear response from a
HTS film at liquid nitrogen temperature. The third-harmonic
power was measured as a function of temperature, input power and
external dc magnetic field for superconducting films. It
argues that the origin of the observed nonlinear response is
likely due to creation of vortices in the film by the microwave
field or an external dc magnetic field.

    The authors are thankful to A. Vorob'ev for HTS films
preparation and to A.A. Andronov for fruitful discussions and
critical comments on the manuscript. This work was  supported by
the Russian Foundation for Basic Research, grant No.
00-02-16158 and partly by grant No. 00-02-16528.

\begin{figure}[!b]
\centering
\leavevmode
\epsfxsize=5.5 cm
{\epsfbox{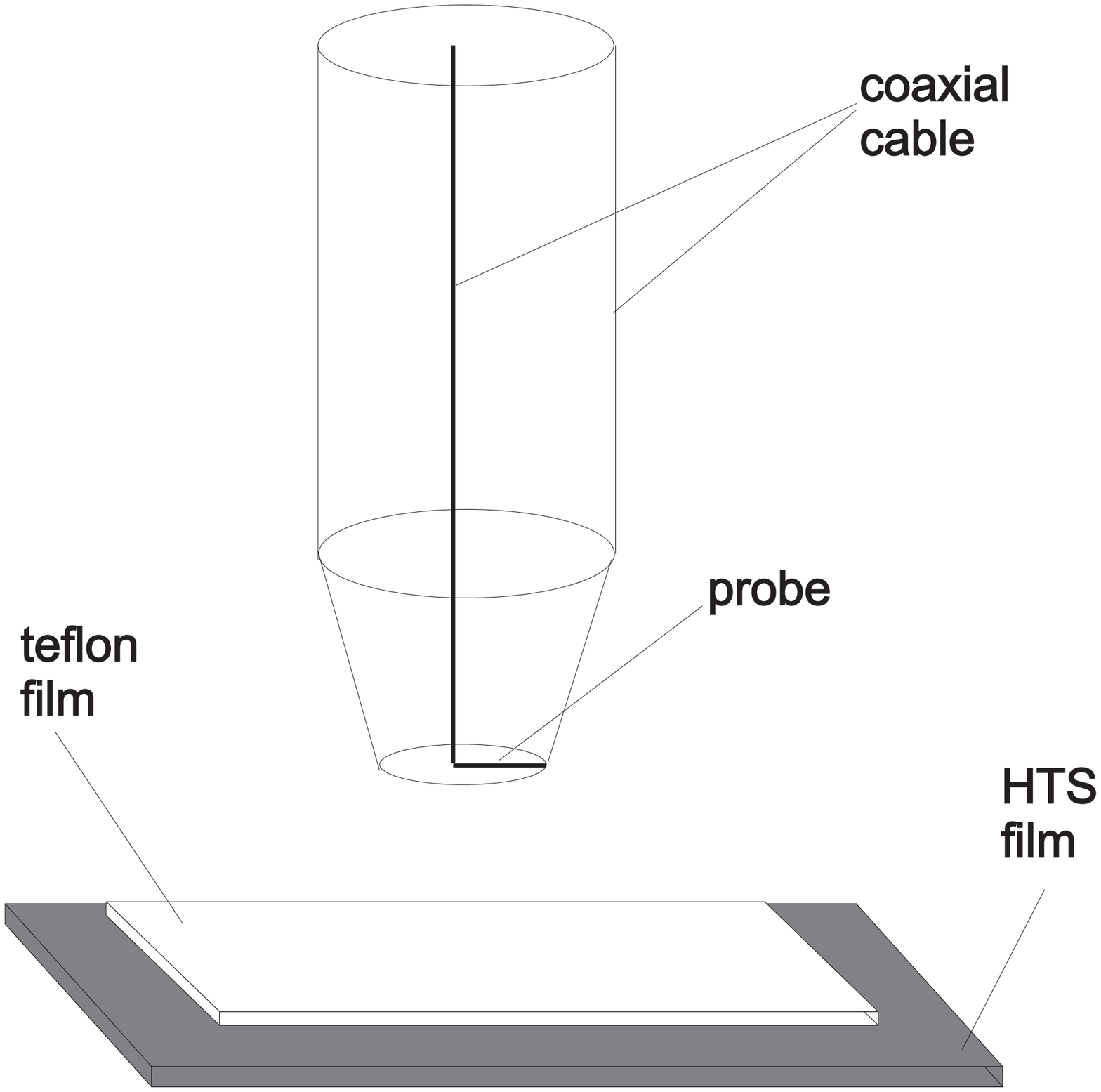}}
\caption{Scheme of the probe.}
\label{zond}
\end{figure}

\begin{figure}[!t]
\centering
\leavevmode
\epsfxsize= 8.5 cm
{\epsfbox{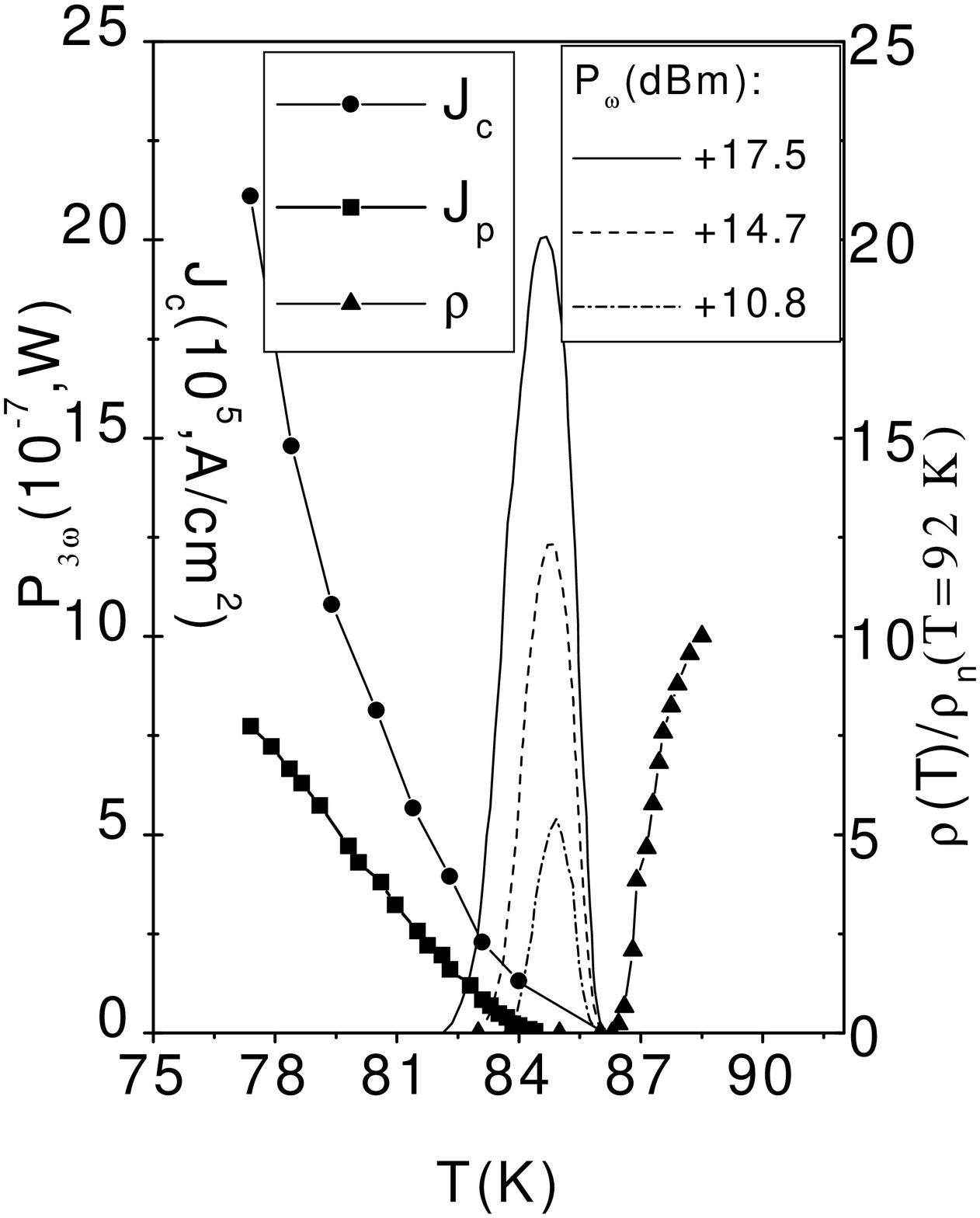}}

\caption{ Temperature
dependences of the third-harmonic signal
 $P_{3\omega}(P_\omega)$ at various input powers, depinning
 current density $J_p$ (squares), current density of vortex penetration
 $J_c$ (circles) and resistivity $\rho$ (triangles) for
 $YBa_2Cu_3O_7$ film.} \end{figure}

\begin{figure}[!b]
\centering
\leavevmode
\epsfxsize=8.5 cm
{\epsfbox{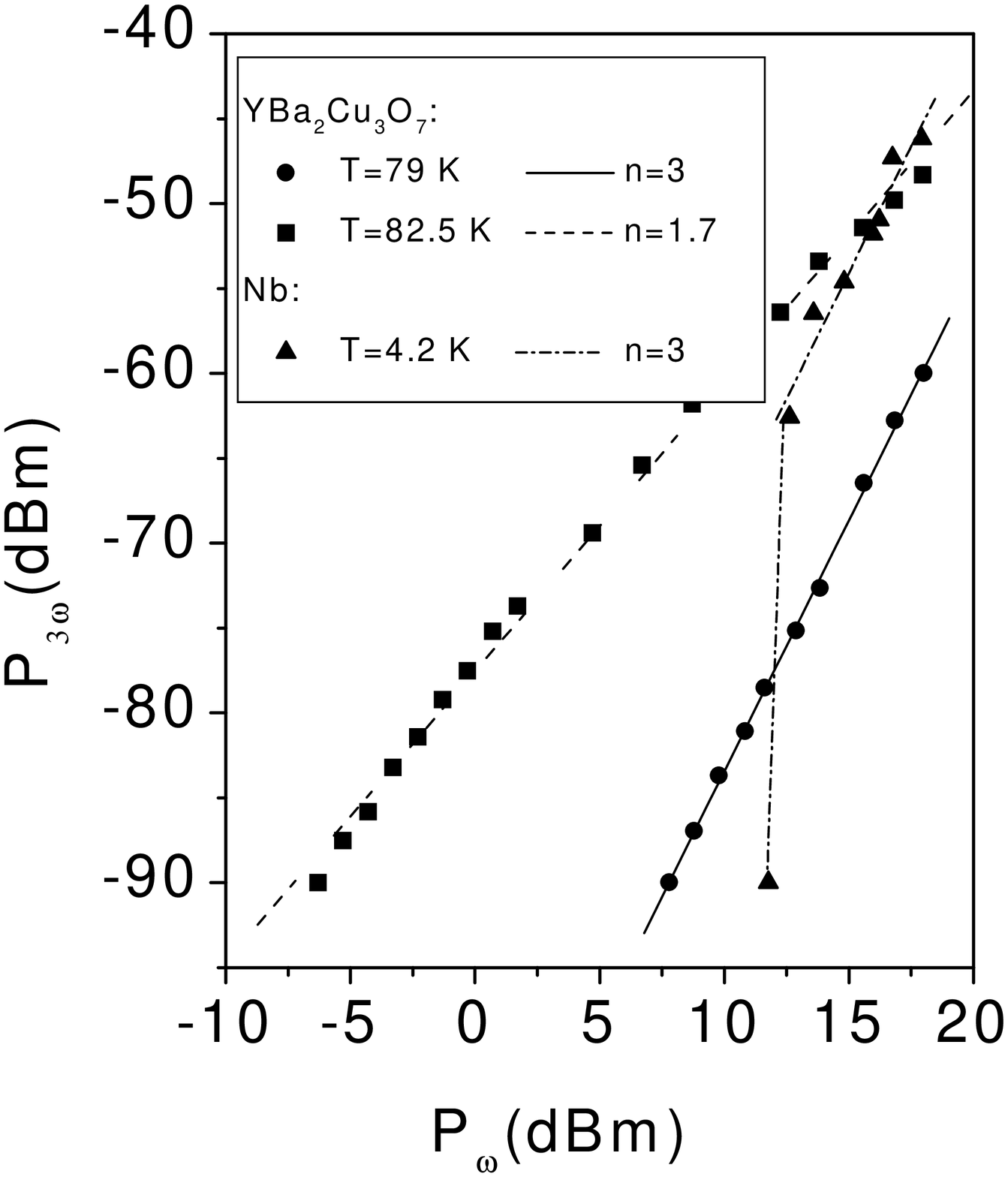}}

\caption{$P_{3\omega}$ vs $P_\omega$ for $YBa_2Cu_3O_7$ film at
temperature near the peak T = 82.5 K (squares), T=77 K
(circles) and Nb film at T= 4.2 K (triangles), fitted by the power-law
$P_{3\omega}\sim A {P_{\omega}}^n$ shown by straight lines.  } \label{a0}
\end{figure}

\begin{figure}[!t]
\centering
\leavevmode
\epsfxsize=8.5 cm
{\epsfbox{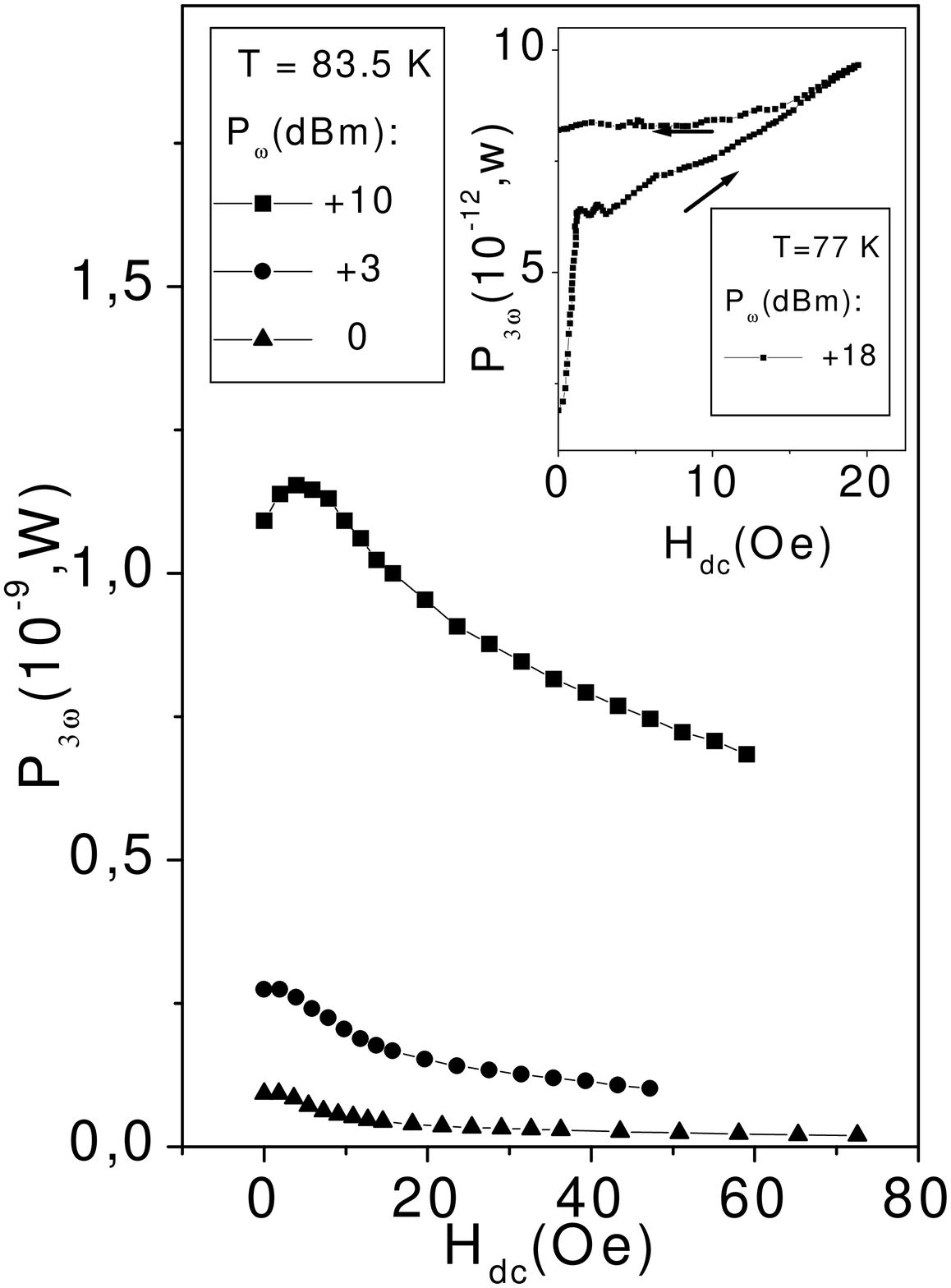}}

\caption{$P_{3\omega}$ vs $H_{dc}$
at different levels of input power at temperature near the peak
of nonlinearity T= 83.5 K for $YBa_2Cu_3O_7$ film.  The inset
shows $P_{3\omega}(H_{dc})$ at temperature T=77 K for $YBa_2Cu_3O_7$
film.  }
\end{figure}

\begin{figure}[!t]
%\centering
%\leavevmode
\epsfxsize=8.5cm {\epsfbox{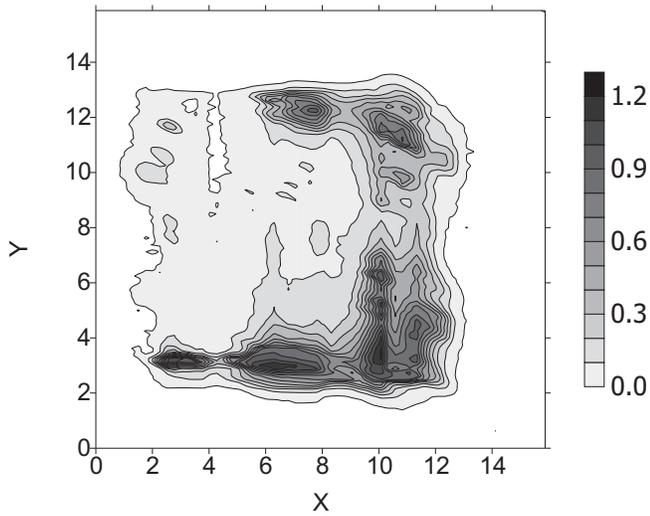}}
\centering
\leavevmode
\caption{The map of  the nonlinear response
of $YBa_2Cu_3O_7$ film at  T=77 K. Sizes of image is shown in
mm. The probe is parallel to the x axis.}
\label{scan}
\end{figure}

\end{document}